\documentstyle [12pt]{article}
\begin{document}
\centerline {\bf PHYSICAL AND COSMOLOGICAL IMPLICATIONS OF A POSSIBLE}
\vskip 3mm
\centerline {\bf CLASS OF PARTICLES
ABLE TO TRAVEL
FASTER THAN LIGHT}
\vskip 15mm
\centerline {\bf L. GONZALEZ-MESTRES}
\vskip 3mm
\centerline {Laboratoire de Physique Corpusculaire, Coll\`ege de France,} 
\centerline {11 pl. Marcellin-Berthelot, 75231 Paris Cedex 05 , France}
\vskip 2mm
\centerline {and}
\vskip 2mm
\centerline {Laboratoire d'Annecy-le-Vieux de Physique des Particules}
\centerline { B.P. 110 , 74941 Annecy-le-Vieux Cedex, France}
\vskip 15mm
{\bf Abstract}
\vskip 2mm
The apparent Lorentz invariance of the laws of physics does not imply that
space-time is indeed minkowskian. Matter made of solutions of Lorentz-invariant
equations would feel a relativistic space-time even if the actual space-time
had a quite different geometry (f.i. a galilean space-time).
A typical example is provided by sine-Gordon solitons in a galilean world.
A "sub-world" restricted to such solitons would be "relativistic", with the 
critical speed of solitons playing the role of the speed of light.
Only the study of the deep structure of matter will unravel the actual geometry 
of space and time, which we expect to be scale-dependent and determined by the 
properties of matter itself.
\par
If Lorentz invariance is only an approximate property of equations describing 
a sector of matter at a given scale, an absolute frame (the "vacuum rest 
frame") may exist without contradicting the minkowskian structure of the 
space-time felt by ordinary particles. But $c$~, the speed of light, will not
necessarily be the only critical speed in vacuum: for instance, superluminal 
sectors of matter may exist related to new degrees of freedom
not yet discovered experimentally. Such particles would not be tachyons:
they may feel different minkowskian space-times with critical speeds much 
higher than $c$ and behave kinematically like ordinary particles apart from the
difference in critical speed. Because of the very high critical speed in
vacuum, superluminal particles will have very large rest energies.
At speed $v > c$ , they are expected to release "Cherenkov" radiation (ordinary
particles) in vacuum. 
\par
We present a discussion of possible physical (theoretical and experimental)
and cosmological implications of such a scenario, assuming that the 
superluminal sectors couple weakly to ordinary matter. The production of 
superluminal particles may yield clean signatures in experiments at very
high energy accelerators. The breaking of Lorentz invariance will be basically
a very high energy and very short distance phenomenon, not incompatible with
the success of standard tests of relativity. Gravitation will undergo 
important modifications when extended to the superluminal sectors. The Big 
Bang scenario, as well as large scale structure, can be strongly influenced 
by the new particles. If superluminal 
particles exist, they could provide most of the cosmic (dark) matter and
produce very high energy cosmic rays compatible with unexplained discoveries 
reported in the literature.
\vskip 2cm
{\bf 1. RELATIVITY, SPACE-TIME AND MATTER}
\vskip 6mm
In textbook special relativity, mikowskian geometry is 
an intrinsic property of space and time: any material body moves
with a universal critical speed $c$ (i.e. at 
speed $v \leq c$), inside a minkowskian space-time governed by Lorentz
transformations and relativistic kinematics. The action itself is 
basically given by
a set of metrics. General relativity includes
gravitation, but the "absoluteness" of the
previous concepts remains even if matter 
modifies the local structure of space and time.
Gravitation is given a geometric description 
within the Minkowskian approach: geometry remains the basic principle of
the theory and provides the ultimate dynamical concept.
This philosophy has widely influenced modern theoretical 
physics and, especially, recent grand unified theories.
\vskip 4mm
On the other hand, a look to various dynamical systems 
studied in the last decades would suggest a more flexible view of
the relation between matter and space-time. 
Lorentz invariance can be viewed as a symmetry of the motion 
equations, in which case no reference to absolute 
properties of space and time is required and the 
properties of matter play the main role. In a two-dimensional 
galilean space-time, 
the equation:
\equation
\alpha ~\partial ^2\phi /\partial t^2~-~\partial ^2\phi /\partial x^2 = F(\phi )
\endequation
with $\alpha$ = $1/c_o^2$ and $c_o$ = critical 
speed, remains unchanged under "Lorentz" transformations leaving 
invariant the squared
interval:
\equation
ds^2 = dx^2 - c_o^2 dt^2
\endequation
so that matter made with solutions of equation (1) 
would feel a relativistic space-time even if the real space-time is actually
galilean and if an absolute rest frame exists in the 
underlying dynamics beyond the wave equation.
A well-known example is provided by the solitons of the sine-Gordon equation, 
obtained taking in (1):
\equation
F(\phi ) = (\omega /c_o)^2~sin~\phi
\endequation
\par 
A two-dimensional universe made of sine-Gordon solitons plunged 
in a galilean world would behave like a two-dimensional minkowskian
world with the laws of special relativity. 
Information on any absolute rest frame would be lost by the solitons.
\vskip 4mm
1-soliton solutions of the sine-Gordon equation are known to 
exhibit "relativistic" particle properties. With $|v|$ $<$ $c_o$ ,
a soliton of speed $v$ is described by the expression:
\equation
\phi _v (x,t) = 
4~arc~tan~[exp~(\pm~\omega c_o^{-1}~(x-vt)~(1- v^2/c_o^2)^{-1/2})]
\endequation
corresponding to a non-dissipative solution with the following properties:
\vskip 3mm
- size $\Delta x = c_o\omega ^{-1}~(1- v^2/c_o^2)^{1/2}$
\vskip 3mm
- proper time $d\tau = dt~(1- v^2/c_o^2)^{1/2}$
\vskip 3mm
- energy $E = E_o~(1- v^2/c_o^2)^{-1/2}$ , $E_o$ being the energy at rest
and $m = E_o/c_o^2$ the "mass" of the soliton
\vskip 3mm
- momentum $p = mv~(1- v^2/c_o^2)^{-1/2}$
\vskip 4mm
\noindent
so that everything looks perfectly "minkowskian" even if the 
basic equation derives from a galilean world with an
absolute rest frame. The actual structure of space and time 
can only be found by going beyond the wave equation
to deeper levels of resolution, similar to the way high 
energy accelerator experiments explore the inner structure of
"elementary" particles.
The answer may then be scale-dependent and matter-dependent.
\vskip 4mm
Free particles move in vacuum, which is known (i.e. from the 
Weinberg-Glashow-Salam theory) to be a material medium
where condensates and other structures can develop. We measure
particles with devices made of particles. We are ourselves made 
of particles, and we are inside the vacuum.
All known particles have indeed a critical speed in 
vacuum equal to the speed of light, $c$ . But a crucial question  remains open:
is $c$ the only critical speed in vacuum, are 
there particles with a critical speed different from that of light?
The question clearly makes sense, as in a 
perfectly transparent crystal it is possible to identify 
at least two critical speeds: the speed of light and
the speed of sound. 
The present paper is devoted to explore a simple nontrivial scenario, with 
several critical speeds in vacuum.
\vskip 6mm
{\bf 2. PARTICLES IN VACUUM}
\vskip 5mm
Free particles in vacuum usually satisfy a dalembertian equation, 
such as the Klein-Gordon equation for scalar particles:
\equation
(c^{-2}~\partial ^2/\partial t^2~-~\Delta)~\phi~~+~~m^2c^2~(h/2\pi )^{-2} 
\phi ~~=~~0
\endequation
where the coefficient of the second time derivative sets $c$ , 
the critical speed 
of the particle in vacuum (speed of light). Given $c$ and the
Planck constant $h$ , the coefficient of the linear term 
in $\phi $ sets $m$ , the mass of the particle.
To build plane wave solutions, we consider the 
following physical quantities given by differential operators:
\vskip 3mm
\centerline {$E = i~(h/2\pi )~\partial /\partial t$~~~~~~,~~~~~~
$\vec{\mathbf p} = -i~(h/2\pi )~\vec{\mathbf \nabla } $}
\vskip 3mm
\noindent
and with the definitions:
\vskip 3mm
\centerline {$x^o = ct$~~~~~~,~~~~~~
$p^o = E/c$
~~~~~~,~~~~~~
$E = (c^2\vec{\mathbf p}^2~+~m^2c^4)^{1/2}$}
\vskip 3mm
\noindent 
the plane wave is given by:
\equation
\phi (x,t) = exp~[-(2\pi i/h)~(p^ox^o-\vec{\mathbf p} . \vec{\mathbf x} )]
\endequation
\noindent
from which we can build position and speed operators [1]:
\equation
\vec{\mathbf x}_{op} = 
(ih/2\pi )~(\vec{\mathbf \nabla 
}_p~-~1/2~(\vec{\mathbf p}^2+m^2c^2)^{-1}\vec{\mathbf p})
\endequation
\noindent
in momentum space, and:
\equation
\vec{\mathbf v} = d\vec{\mathbf x}_{op} /dt = (2\pi i/h)~[H ,
\vec{\mathbf x}_{op}] = (c/p_o)~\vec{\mathbf p}
\endequation
\noindent
where $H$ is the hamiltonian and the brackets mean commutation. We then get:
\vskip 3mm
\centerline {$p_o = mc~(1~-~v^2/c^2)^{-1/2}~~~~~~,~~~~~~\vec{\mathbf p} = 
m\vec{\mathbf v}~(1~-~v^2/c^2)^{-1/2}$}
\vskip 3mm
\noindent
and, at small $v/c$ :
\vskip 3mm
\centerline {$E_{free} \simeq 1/2~mv^2 ~~~~~~,~~~~~~\vec{\mathbf p} 
\simeq m\vec{\mathbf v}~~~~~,~~~~~\vec{\mathbf x}_{op}~\simeq
~i~(h /2\pi) ~\vec {\mathbf \nabla} _p$}
\vskip 3mm
\noindent
in which limit, taking $H = 1/2~mv^2~+~V(\vec{\mathbf x}_{op})$ , 
we can commute $H$ et $\vec {\mathbf v}$ and obtain:
\vskip 3mm
\centerline {$\vec{\mathbf F} = - \vec{\mathbf \nabla }V = 
m~d\vec{\mathbf v} /dt$}
\vskip 3mm
\noindent
which shows that $m$ is indeed the inertial mass.
\vskip 4mm
Superluminal sectors of matter can be consistently 
generated, with the conservative choice of leaving the Planck constant
unchanged, replacing in the above construction the 
speed of light $c$ by a new critical speed $c_i$ $>$ $c$ 
(the subscript $i$ stands for the $i$-th superluminal sector). All
previous concepts and
formulas remain correct, leading to particles with 
positive mass and energy which are not tachyons
and have nothing to do with previous proposals in this field [2]. 
For inertial mass $m$ and critical speed $c_i$ , the new particles
will have rest energies:
\equation 
E_{rest} = mc_i^2
\endequation
To produce superluminal mass at accelerators 
may therefore require very large energies.
In the "non-relativistic" limit $v/c_i$ $\ll $ 1 , kinetic 
energy and momentum will remain given by the same 
non-relativistic expressions as before.
Energy and momentum conservation will in principle not be 
spoiled by the existence of several critical speeds in vacuum:
conservation laws will as usual hold for phenomena leaving the vacuum unchanged.
\vskip 6mm
{\bf 3. A SCENARIO WITH SEVERAL CRITICAL SPEEDS IN VACUUM}
\vskip 5mm
Assume a simple and schematic scenario, with several sectors of matter:
\vskip 3mm
- the "ordinary sector", made of "ordinary particles" with a 
critical speed equal to the speed of light $c$ ;
\vskip 3mm
- one or more superluminal sectors, where particles have 
critical speeds $c_i$ $\gg $ $c$ in vacuum, 
and each sector is assumed to have its own Lorentz invariance 
with $c_i$ defining the metric.
\vskip 4mm
Several basic questions arise: can different sectors 
interact, and how? what would be the conceptual
and experimental consequences? can we observe the superluminal 
sectors and detect their particles? what would be the best
experimental approach? It is obviously impossible to give 
general answers independent of the details of the scenario
(couplings, symmetries, parameters...), 
but some properties and potentialities can be pointed out.
\vskip 5mm
{\bf 3a. Lorentz invariance(s)}
\vskip 4mm
Even if each sector has its own "Lorentz invariance" involving 
as the basic parameter the critical speed in vacuum of
its own particles, interactions between two different 
sectors will break both Lorentz invariances. With an interaction
mediated by 
complex scalar fields preserving apparent Lorentz 
invariance in the lagrangian density (e.g. with a $|\phi _o(x)|^2|\phi _1(x)|^2$
term where $\phi _o$ belongs to the ordinary sector and $\phi _1$
to a superluminal one), the Fourier expansion of the scalar fields shows 
the unavoidable breaking of Lorentz invariances.
The concept of mass, as a relativistic invariant, becomes equally approximate 
and sectorial.
\vskip 3mm
 Even before considering interaction between different sectors, Lorentz 
invariance
for all sectors simultaneously will at best be explicit in a single
inertial frame (the $vacuum$ $rest$ $frame$, i.e. the "absolute" rest 
frame). Then, apart from space rotations,
no linear space-time transformation can simultaneously preserve 
the invariance of lagrangian densities for two different sectors.
However, it will be impossible to identify the vacuum rest 
frame if only one sector produces measurable effects
(i.e. if superluminal particles and their influence on the ordinary 
sector cannot be observed).
In our approach, the Michelson-Morley result is not incompatible 
with the existence of some "ether" as
suggested by recent results in particle physics: if 
the vacuum is a material medium where fields 
and order parameters can condense, it may well
have a local rest frame. If superluminal particles couple
weakly to ordinary matter, their effect on the ordinary sector will occur at
very high energy and short distance, far from
the domain of successful
conventional tests of Lorentz invariance. Nuclear and 
particle physics experiments may open new windows in this field.
Finding some track of a superluminal sector (e.g. through 
violations of Lorentz invariance in the ordinary sector) may
be the only way to experimentally discover the vacuum rest frame.
\vskip 5mm
{\bf 3b. Space, time and supersymmetry}
\vskip 4mm
If the standard minkowskian space-time is not a compulsory 
framework, we can conceive fundamentally different descriptions
of space and time. Just to give examples, we 
can consider three particular scenarios:
\vskip 4mm
{\bf Galilean case.} There is no absolute critical speed for 
particles in vacuum, and space-time transforms according
to galilean transformations. The sectorial critical speeds are then the analog
of the critical speed of the solitons in the above-mentioned sine-Gordon system
sitting in a galilean world. 
\vskip 4mm
{\bf Minkowskian case.}
There is an absolute critical speed $C$ in vacuum, 
with $C\gg c$ and $C \gg c_i $ , generated from a
cosmic Lorentz invariance not yet found experimentally. As long as physics
happens at speed scales much lower than $C$ , the situation is analog to
the galilean case. 
Particles with critical speed equal to $C$ will most likely be
weakly coupled to particles from the ordinary and superluminal sectors and 
be produced only at very high energy. However, massless particles of this type 
may play a role in low energy phenomena (e.g. a cosmic gravitational
interaction). We would reasonably expect
particles from the sector with critical speed $C$ (the "cosmic"
sector) to be the actual constituents of matter. 
\vskip 4mm
{\bf Spinorial case.}
Since spin-1/2 particles exist in nature, and they do not
form representations of the rotation group $SO(3)$ but rather of its covering
group $SU(2)$ , it seems reasonable to attempt a spinorial description of
space-time. Lorentz invariance is not required for that purpose, as the basic
problem already exists in non-relativistic quantum mechanics.
A simple way to relate space and time to a spinor $C^2$ complex two-dimensional
space would be, for a spinor $\xi $ with complex coordinates $\xi _1$ and $\xi
_2$ , to identify time with the spinor modulus. At first sight, this has the 
drawback of positive-definiteness, but it naturally sets an arrow of time
as well as an origin of the Universe.
Then, $t$ = $| \xi |$ could be a cosmic time for an expanding Universe
where space would be parameterized by $SU(2)$ transformations and, locally, by 
its generators which form a vector representation of $SU(2)$ (the tangent
space to the $S^3$ hypersphere in $C^2$ made topologically 
equivalent to $R^4$). The $SU(2)$ invariant metrics 
$| d\xi |^2$ = $| d\xi _1 | ^2 $ $+$ $| d\xi _2| ^2 $ sets 
a natural relation between local
space and time units. However, the relevant 
speed scale does not a priori correspond
to any critical speed for particles in vacuum. Furthermore, the physically
relevant local space and time scales will be determined only by the dynamical
properties of vacuum. Radial straight lines in the $R^4$ space, starting from
the point $\xi = 0$ , may naturally define the vacuum rest frame at cosmic
scale. Any real fonction defined on the complex manifold $C^2$ will be a
fonction of $\xi $ and $\xi ^{\ast}$ . Spin-1/2 fields will correspond to
linear terms in the components of $\xi $ and $\xi ^{\ast}$ . 
\vskip 4mm
Independently of the critical speed, spin-$1/2$ particles in the 
vacuum rest frame can have a well-defined helicity:
\equation 
(\vec \sigma .\vec {\mathbf p })~\mid \psi >~=~\pm p~\mid \psi >
\endequation
which, for free massless particles with critical 
speed $c_i$ , leads to the Weyl equation:
\equation
(\vec \sigma .\vec {\mathbf p })~\mid \psi >~=~\pm ~(E/c_i)~\mid \psi >
\endequation
remaining invariant under sectorial Lorentz transformations 
with critical speed $c_i$ .
When the $SU(2)$ group acting on spinors
is dynamically extended to a sectorial Lorentz 
group, massless helicity eigenstates form irreducible
representations of the sectorial Lorentz Lie algebra which, 
when complexified, can be split into "left"
and "right" components. While the original $SU(2)$
symmetry can be a fundamental property of space and time, 
the Lorentz group and its chiral components
are not fundamental symmetries.
\vskip 4mm
In the spinorial space-time,
we may attempt to relate the $N=1$ supersymmetry generators 
to the spinorial momenta $\partial /\partial \xi _{\alpha } $
$(\alpha ~=~1~,~2)$ and to their hermitic conjugates.
Extended supersymmetry could similarly be linked to a set of 
spinorial coordinates 
$\xi _{\alpha }^j$ $(j~=~1~,...,~N~;~\alpha~=~1~,~2)$ and 
to their hermitic conjugates,
the index $j$ being the internal symmetry index.
Contrary to ordinary superspace [4], the new spinorial coordinates would not 
be independent from space-time coordinates.
Time can be the modulus of the $SU(2)\otimes SO(N)$ spinor, taking: 
$t^2 ~=~\Sigma_{j=1}^N \Sigma_ {\alpha = 1}^2 \mid \xi _\alpha ^j \mid^2$ , 
and as before the three space coordinates would correspond to the
directions in the $SU(2)$ tangent space.
Then, the $\xi ^j_{\alpha }$ would be "absolute" cosmic spinorial coordinates. 
For each sector with critical speed $c_i$ ,
an approximate sectorial supersymmetry can be dynamically generated,
involving "left" and "right" chiral spaces and
compatible with the sectorial Lorentz invariance.
\vskip 4mm
These examples suggest that, in abandoning 
the absoluteness of Lorentz invariance, 
we do not necessarily get a poorer theory.
New interesting possibilities appear in the domain of fundamental symmetries
as a counterpart to the abandon of a universal Lorentz group.
\vskip 5mm
{\bf 3c. Gravitation}
\vskip 4mm
Mass mixing between particles from different dynamical sectors may occur. 
Although we expect such phenomena to be weak, they could be
more significant for very light particles. 
Because of mixing between different sectors, mass ceases 
to be a Lorentz-invariant
parameter. 
\vskip 4mm
Gravitation is a gauge interaction related to invariance 
under local linear transformations of space-time.
The graviton is a massless ordinary particle, propagating 
at $v=c$ and associated to ordinary Lorentz invariance. Therefore,
it is not expected to play a universal role in the presence of superluminal
sectors. In a supersymmetric scheme, it will belong to a 
sectorial supermultiplet of ordinary particles (supergravity).
\vskip 4mm
Gravitational coupling of superluminal particles to ordinary ones is 
expected to be
weak. Assuming that each superluminal sector
has its own Lorentz metric
$g_{[i] \mu \nu }$
($[i]$ for the $i$-th sector), with $c_i$ setting the speed scale, 
we may expect each sector to generate
its own gravity with a coupling constant $\kappa _i$ and a new 
sectorial graviton traveling at speed $c_i$ .
In a sectorial supersymmetric (supergravity) theory,
each sectorial graviton may belong to a superluminal 
supermultiplet with critical speed $c_i$ .
Gravitation would in all cases be a single and universal interaction only in the
limit where all $c_i$ tend to $c$
and where a single metric, as well as possibly 
a single and conserved supersymmetry, can be used.
\vskip 4mm
As an ansatz, we can assume that the static gravitational coupling 
between two different sectors
is lowered by a factor proportional to a positive power of the 
ratio between the two critical speeds
(the smallest speed divided by the largest one). 
Static gravitational forces between ordinary matter 
and matter of the $i$-th
superluminal sector would then be proportional to a positive power of 
$c/c_i$ which can be a very small number.
"Gravitational" interactions between two sectors
(including "graviton" mixing) can 
be generated through the above considered pair of complex scalar fields,
although this will lead to anomalies in "gravitational" forces for both sectors.
In any case, it seems that concepts
so far considered as very fundamental (i.e. the universality of 
the exact equivalence between inertial and gravitational mass)
will now become approximate sectorial properties
(like the concept of mass itself), even if the 
real situation may be very difficult to
unravel experimentally. Gravitational properties of 
vacuum are basically unknown in the new scenario.
\vskip 5mm
{\bf  3d. Other dynamical properties}
\vskip 5mm
No basic consideration (apart from Lorentz invariance,
which is not a fundamental symmetry in the present 
approach) seems to prevent "ordinary" interactions other than gravitation from
coupling to the new dynamical sectors with their usual strengths. 
Conversely, ordinary particles can in 
principle couple to interactions mediated by superluminal objects.
In the vacuum rest frame, covariant derivatives can
be written down for all particles 
and gauge bosons independently of their critical speed in vacuum. 
Field quantization is performed in hamiltonian formalism, which 
does not require explicit Lorentz invariance, and quantum
field theory can use non-relativistic gauges. 
We do not expect fundamental consistency problems
from the lack of Lorentz invariance, which in quantum 
field theory is more a physical requirement than a real need,
but experiment seems to
suggest that superluminal particles have 
very large rest energies or couple very weakly to the ordinary sector.
\vskip 4mm
Stability under radiative corrections (e.g. 
of the existence of well-defined "ordinary" and "superluminal" sectors)
is not always ensured. As the critical speed 
is related to particle properties in the region of very 
high energy and momentum,
the ultraviolet behaviour of the renormalized theory 
(e.g. renormalized propagators) will be crucial.
However, work on supersymmetry, supergravity and other theories
suggests that technical solutions can be found to 
preserve the identity of each sector
as well as the stability of the scheme.
Although it seems normal to assume that the superluminal sector 
is protected by a quantum number and that the
"lightest superluminal particle" will be stable, 
this is not unavoidable and we may be inside a sea of very long-lived
superluminal particles which decay into ordinary particles 
and/or into "lighter" (i.e. with lower rest energies) superluminal ones.
\vskip 4mm
Finally, it should be noticed that we have kept the value of the
Planck constant unchanged when building the superluminal sectors.
This is not really arbitrary, as conservation and 
quantization of angular momentum make it natural
if the superluminal sectors and the ordinary sector interact.
Strictly speaking, $h$ does not play any fundamental 
dynamical role in the discussion of {\bf Section 2} and its use
at this stage amounts to setting an overall scale.
It seems justified to start the study of superluminal 
particles assuming that their quantum properties
are not different from those of ordinary particles.
\vskip 5mm
{\bf 3e. Some signatures}
\vskip 4mm
If superluminal particles couple to ordinary matter, 
they will not often be found traveling at a speed $v > c$ (except
near very high-energy accelerators where they can be produced,
or in specific astrophysical situations). At 
superluminal speed, they are expected to release "Cherenkov"
radiation, i.e. ordinary particles whose emission in 
vacuum is kinematically allowed
or particles of the $i$-th superluminal
sector for $v>c_i$. Thus, superluminal particles will eventually be decelerated
to a speed $v \leq c$~.
The nature and rate of "Cherenkov" radiation in 
vacuum will depend on the superluminal particle and can be very weak in
some cases.
Theoretical studies of tachyons rejected [3] the 
possibility of "Cherenkov" radiation in vacuum because tachyons are not really
different from ordinary particles (they sit in a different 
kinematical branch, but are the same kind of matter). In our case,
we are dealing with a different kind of matter 
but superluminal particles will always be in the region
of $E$ and $\vec{\mathbf p}$ real, $E$ $=$ $(c_i\vec{\mathbf p}^2~+~m^2c_i^4)^{1/2} >
0$ , and can emit "Cherenkov" radiation.
\vskip 4mm
In accelerator experiments, this "Cherenkov" 
radiation may provide a clean signature allowing to identify
some of the produced
superluminal particles (those with the strongest "Cherenkov" effect).
Other superluminal particles may couple 
so weakly to ordinary matter that "Cherenkov" deceleration in vacuum
occurs only at large astrophysical distance scales.
If this is the case,
one may even, for the far future, think of a very high-energy
collider as the device to emit modulated and directional superluminal signals.
\vskip 4mm
Hadron colliders (e.g. LHC) are in principle the safest way to possibly produce
superluminal particles, as quarks couple to all known interactions.
$e^+e^-$
collisions should be preferred only if superluminal particles 
couple to the electroweak sector. In an accelerator experiment, 
a pair of superluminal particles with inertial mass $m$ would be produced at $E$
(available energy) $=~2mc_i^2$ and 
Cherenkov effect in vacuum will start slightly
above, at $E~=~2mc_i^2~+~mc^2$ $=$ $2mc_i^2 (1~+~1/2~c^2/c_i^2)$
$\simeq $ $2mc_i^2$ .
The Cherenkov cones will quickly become broad, leading 
to "almost $4\pi $" events in the rest frame of the superluminal pair.
\vskip 4mm
Apart from accelerator
experiments, the search for abnormal effects in 
low energy nuclear physics, electrodynamics 
and neutrino physics (with neutrinos moving close
to speed of light with respect to the vacuum rest frame) deserves consideration.
Mixing with superluminal sectors is likely to 
produce the strongest effects on photons and light neutrinos.
In underground and underwater experiments, dark matter superluminal particles 
(see {\bf Section 4}) can produce electron, nucleon or nucleus recoil
but also inelastic events (particularly interesting 
if there is no conserved quantum number protecting the relevant superluminal
sector). High-energy cosmic ray events can yield crucial 
signatures (see {\bf Section 5} ).
\vskip 6mm
{\bf 4. COSMOLOGICAL IMPLICATIONS}
\vskip 5mm
Superluminal particles may have played a cosmological 
role leading to substantial changes in the
"Big Bang" theory and to a reformulation of several fundamental 
problems (Planck limit, cosmological constant,
horizon, inflation, large scale structure, dark matter...).
They may be leading the present evolution of the Universe at 
very large scale.
\vskip 5mm
{\bf 4a. The Universe}
\vskip 5mm
If superluminal sectors exist and Lorentz invariance is only an approximate
sectorial property, the Big Bang scenario may become quite different, as:
a) Friedmann equations do no longer govern the global 
evolution of the Universe, which will be influenced
by new sectors of matter coupled to new forces 
and with different couplings to gravitation; b) gravitation itself
will be modified, and can even disappear, at distance 
scales where Lorentz invariance does no longer hold;
c) at these scales, conventional extrapolations to a "Big 
Bang limit" from low energy scales do not make sense;
d) because of the degrees of freedom linked to superluminal 
sectors, the behaviour of vacuum will be different from standard
cosmology; e) the speed of light is no longer an upper 
limit to the speed of matter.
If real space and time are not the basic space-time coordinates 
(e.g. if space-time is actually described by the above proposed
spinorial coordinates transforming under a $SU(2)$ group), 
the usual Big Bang framework may become 
unadapted below certain space and time scales,
even if a $t=0$ limit exists.
In the above spinorial space-time, naturally providing an expanding Universe, 
the Big Bang limit can be 
identified to the point $\xi = 0$ . The mathematical relation between the 
physical, local time scale and the cosmic time $t = ~\mid \xi \mid $ would
depend on dynamics. 
\vskip 4mm
Each sectorial Lorentz invariance is expected to 
break down below a critical distance scale, when the Lorentz-invariant
equations (and Lorentz-covariant degrees of freedom) cannot be used 
and a new dynamics appears.
For a sector with critical speed $c_i$ and apparent Lorentz 
invariance at distance
scales larger than $k_i^{-1}$ , where $k_i$ is a critical wave 
vector scale, we can expect
the appearance of a critical temperature $T_i$
given approximately by:
\begin{equation}
k~T_i~\approx ~\hbar ~c_i~k_i
\end{equation}
where $k$ is the Boltzmann constant and $\hbar $ the Planck constant. 
Above $T_i$ , the vacuum will not necessarily allow for 
the previously mentioned particles
of the $i$-th sector and new forms of matter can appear. If $k_0$ stands for the
critical wave vector scale of the ordinary sector,
above $T_0$ $\approx $ $k^{-1}\hbar ck_0$
the Universe may 
have contained only superluminal particles 
whereas superluminal and ordinary
particles coexist below $T_0$ . It may happen that some ordinary
particles exist above $T_0$ , but with different
properties (like sound above the melting point).
\vskip 4mm
If Lorentz invariance is not an absolute property of space-time, 
ordinary particles did most likely not govern the beginning of
the Big Bang (assuming such a limit exists)
and dynamical correlations have been able to propagate
must faster than light
in the very early Universe.
The existence of superluminal particles, and of the vacuum degrees 
of freedom which generate such excitations,  
seems potentially able
to invalidate arguments leading to the
so-called "horizon problem" and "monopole problem", because:
a) above $T_0$ , particles and dynamical correlations are 
expected to propagate mainly at superluminal speed,
invalidating conventional estimates of the "horizon size";
b) below $T_0$ , annihilation and decays of superluminal
particles into ordinary ones will release very large
amounts of kinetic energy from the rest 
masses ($E=mc_i^2$ , $c_i \gg c$) and generate
a fast expansion of the Universe. Conventional inflationary 
models rely on Friedmann equations and will not
hold in the new scenario. New inflationary models can 
be considered, but their need is far from obvious.
If a "Big Bang" limit exists,
and if
a generalized form of Friedmann equations can be written down 
incorporating all sectors and forces,
a definition of the horizon distance would be:
\begin{equation}
d_H(t)~=~R(t)~\int_0^t ~C(t') R(t')^{-1}~dt'
\end{equation}
where $d_H$ is the generalized horizon distance, $R(t)$ is 
the time-dependent cosmic scale factor and $C$ is the maximum of all
critical speeds. 
This definition is realistic even for ordinary particles, which can 
be radiated by superluminal ones or produced by
their annihilation.
$C$ can be infinite if one of the sectors has no critical speed 
(as in the usual galilean space-time), or
if an infinite number of superluminal sectors exist.
The homogeneity and isotropy of the present Universe, as manifested 
through COBE results, are now natural properties.
\vskip 5mm
{\bf 4b. Particles in the Universe}
\vskip 5mm
The coupling between the ordinary sector and the superluminal ones
will influence black hole dynamics.
A detectable flux of magnetic monopoles (which can
be superluminal) is not excluded, as the 
"horizon problem" can be eliminated without the standard inflationary scheme.
Long range correlations introduced by superluminal 
degrees of freedom can play a role in the formation of
objects (i.e. strings) leading to large scale structure of the Universe.
\vskip 4mm
Physics at grand unified scales can present new interesting features.
Grand unified symmetries
are possible as sectorial symmetries of the 
vacuum degrees of freedom, even if ordinary particles do not exist
above $T_0$ . But analytic extrapolations 
(e.g. of running coupling constants) cannot be performed
above the phase transition temperature. 
If $kT_0$ is not higher
than $\approx $ 10$^{14}$ $GeV$
($k_0^{-1}$ $\approx $ 10$^{-27}$ $cm$ , time scale
$\approx $ 10$^{-38}$ $s$), the formation of a symmetry-breaking
condensate in vacuum 
may have occurred above $T_0$ and remain below the transition temperature.
Because of
superluminal degrees of freedom and of phase
transitions at $T_0$ and at all $T_i$ ,
it seems impossible to set a "natural time scale"
based on extrapolations from the low energy sector
(e.g. at the Planck time $t_p$ $\approx $ 10$^{-44}$ $s$ 
from Newton's constant).
Arguments leading to the "flatness" or "naturalness"
problem, as well as
the concept of the cosmological constant and 
the relation between critical density and Hubble's "constant" (one of the basic
arguments for ordinary dark matter at cosmic scale),
should
be reconsidered.
\vskip 4mm
At lower temperatures, superluminal particles do not 
disappear. In spite of obvious limitations coming from
annihilation, possible decays, decoupling and "Cherenkov radiation"
(although such phenomena have by themselves cosmological implications),
they can produce important effects in the evolution of the Universe.
Superluminal matter may presently be dark, with an unknown 
coupling to gravitation and coupled to
ordinary matter by new, unknown forces.
Concentrations of superluminal matter would not necessarily follow
the same pattern as "ordinary" galaxies and clusters of
galaxies (made of ordinary matter), nor would they need to occur at 
the same places. Although we
expect correlations between the distributions of ordinary and 
superluminal matter at large scales, it may be difficult
(but not impossible, e.g. in the presence of coupled 
gravitational singularities involving several sectors)
to reasonably assume that the gravitational role of galactic halos 
be due to superluminal
particles.
Superluminal particles can be very abundant and even provide most 
of the (dark) matter at cosmic scale,
but they will be extremely difficult
to detect if they interact very weakly with ordinary matter. 
Similarly, astrophysical objects made of superluminal matter may
elude all conventional observational techniques
and be extremely difficult to find. 
\vskip 5mm
{\bf 4c. Possible signatures}
\vskip 5mm
If superluminal particles are very abundant,
they can, in spite of their expected weak coupling to gravitation,
produce some observable gravitational effects. It 
is not obvious how to identify the superluminal origin of a collective
gravitational phenomenon, but clean signatures may exist 
in some cases (e.g. in gravitational collapses
or if it were possible to detect superluminal gravitational waves).
If astrophysical concentrations of superluminal matter 
produce high-energy particles (ordinary or superluminal),
cosmic rays may provide a unique way to detect such objects 
(see {\bf Section 5}).
Direct detection of particles from superluminal 
matter around us in underground and underwater detectors cannot be discarded:
\vskip 4mm
- If $m$ , $v$ and $p$ are the mass, speed 
and momentum of a relativistic superluminal particle (i.e. at $v~\simeq ~c_i$),
and $M$ the mass of an ordinary particle (electron, proton, neutron...) from 
the target, we expect recoil momenta
$p_R~\sim ~p$ and elastic recoil energies $E_R~\sim ~p~c$ .
For $m~\sim ~1~GeV/c^2$ , $v~\simeq ~c_i~\sim ~10^6~c$ 
and $\gamma ~=~(1~-~v^2/c_i^2)^{-1/2}~\simeq ~10^6$ , recoil energies
$E_R~\sim ~10^{12}~GeV$ are obtained similar to the highest-energy 
cosmic rays observed.
Inelastic events can produce much higher energies, 
comparable to that of the incoming superluminal particle
($\sim ~10^{18}~GeV$ in the previous example).
\vskip 4mm
- Low-energy superluminal particles may also produce detectable events. For 
instance, at $v~\simeq ~c$ (after 
"Cherenkov" deceleration
in vacuum) or at $v~\sim ~v_e~<~c$ 
($v_e$ being some local escape velocity $\gg ~10^{-3}~c$), a superluminal 
particle can produce
recoil protons and neutrons with  energies $\gg ~1~keV$ , up 
to $E_R~\sim ~1~GeV$ (for $m~\sim ~1~GeV/c^2$ and $v~\sim ~c$).
The comparatively high energy of such events can make then detectable
and identifiable, even at very small rates, especially in
very large volume detectors (e.g. the Cherenkov detectors for neutrino 
astronomy [5]).
Again, inelastic events can yield much higher energy deposition.
\vskip 4mm
- In cryogenic detectors, recoil spectra with average 
energies $E_R~\gg ~10^{-6}~Mc^2$ ($M~=$ mass of the target nucleus)
can be, according to standard halo models,
a signature for particles with escape velocity $v_e~\gg ~10^{-3}c$ 
not obeying the usual gravitational laws.
Other signatures for particles with escape velocity substantially 
different from $10^{-3}c$ may come from 
comparison between results obtained with two different targets 
(e.g. a hydrogen target and a iodine, xenon or tungsten target).
Such particles, whose gravitational behaviour would not fit with conventional  
halo models, would possibly belong to the superluminal sectors.
\vskip 6mm
{\bf 5. SUPERLUMINAL PARTICLES AND HIGH-ENERGY COSMIC RAYS}
\vskip 5mm
Annihilation of pairs of superluminal particles into ordinary ones can
release very large kinetic energies and provide a new
source of high-energy cosmic rays. If 
some superluminal sectors are not protected by a conserved quantum number,
their decays may play a similar role.
Cosmic rays, especially at high energy and not coming from conventional
sources, can therefore be crucial to find a track of superluminal matter.
\vskip 5mm
{\bf 5a. Ordinary primaries} 
\vskip 4mm 
$Annihilation$ of pairs of slow
superluminal particles
$into$ $ordinary$ $particles$ (and similarly, decays), 
releasing very high kinetic energies
from the superluminal rest energies (the relation $E~=~mc_i^2$),
would yield a unique $cosmic$ $signature$ allowing cosmic ray 
detectors to search
for this new kind of matter in the present Universe. 
$Collisions$ (especially, inelastic with very large energy 
transfer) of high-energy superluminal particles $with$
$extra$-$terrestrial$ $ordinary$ $matter$ may also yield high-energy 
ordinary cosmic rays.
High-energy superluminal particles can be produced from acceleration, 
decays, explosions... in astrophysical objects made of
superluminal matter. Pairs of slow superluminal particles can also 
annihilate into particles of another superluminal sector
with lower $c_i$ , converting most of the rest energies into a 
large amount of kinetic energy.
Superluminal particles moving at $v>c$ can release anywhere $"Cherenkov"$
$radiation$ $in$ $vacuum$, i.e.
 spontaneous emission of particles of a
lower critical speed $c_i$ (for $v>c_i$) including ordinary ones,
providing a new source of (superluminal or ordinary) 
high-energy cosmic rays.
\vskip 5mm
{\bf 5b. Superluminal primaries} 
\vskip 4mm
High-energy superluminal particles
can directly $reach$ $the$ $earth$ and undergo
collisions inside the atmosphere, producing many secondaries like
ordinary cosmic rays. They can also interact with the rock or
with water near some underground or underwater detector,
coming from the atmosphere or after having crossed the earth, 
and producing clear signatures.
Contrary to neutrinos, whose flux is strongly attenuated by the 
earth at
energies above $10^6$ $GeV$ , superluminal particles will in
principle not be stopped by earth at these energies.
In inelastic collisions, high-energy superluminal primaries can 
transfer most of their energy to ordinary particles.
Even with a very weak interaction probability, 
and assuming that the superluminal primary does not produce any ionization,
the rate for superluminal cosmic ray events can be observable
if we are surrounded by important concentrations of superluminal
matter. Background rejection would be further enhanced by 
atypical ionization properties.
\vskip 5mm
{\bf 5c. Event interpretation}
\vskip 4mm
The possibility that superluminal matter exists, and that it plays 
nowadays an important role in our Universe,
should be kept in mind when addressing the two basic questions 
raised by the analysis of any cosmic ray event:
a) the nature and properties of the cosmic ray primary; b) 
the identification (nature and position) of
the source of the cosmic ray. 
\vskip 4mm
If the primary
is a superluminal particle, it will escape conventional criteria 
for particle identification
and most likely produce a specific signature 
(e.g. in inelastic collisions) different from those of ordinary 
primaries (see also {\bf Subsection 4c}).
Like neutrino events, in the absence of 
ionization (which will in any case be very weak) we 
may expect the event to start anywhere inside the detector.
Unlike very high-energy neutrino events, 
events created by superluminal primaries can 
originate from a particle having crossed the earth.
An incoming, relativistic superluminal particle with momentum $p$ and
energy $E_{in} \simeq p~ c_i $ , hitting an ordinary particle at rest,
can, for instance, release most of its energy into 
two ordinary particles with 
momenta close to $p_{max}~=~1/2~p~c_i~c^{-1}$ and oriented back
to back in such a way
that the two momenta almost cancel. Then, an energy $E_R \simeq E_{in} $
would be transferred to ordinary secondaries. At very high energy, such events
would be easy to identify in large volume detectors, even at very small rate.
\vskip 4mm
If the
source is superluminal, it can be located anywhere 
(and even be a free particle) and will not necessarily be at the same place as
conventional sources
of ordinary cosmic rays. High-energy cosmic ray events 
originating form superluminal sources
will provide hints on the location of such 
sources and be possibly the only way to observe them.
\vskip 4mm
The energy dependence of the events should be taken into account.
At very high energies,
the Greisen-Zatsepin-Kuzmin cut-off [6] does not in principle hold for
cosmic ray events originating from superluminal matter:
this is obvious if the primaries are superluminal particles
that we expect to interact very weakly with the cosmic microwave 
background,
but is also true for ordinary primaries as we do not expect them to
be produced at the locations of ordinary sources and there is no upper 
bound to their energy around $100~EeV$. However, 
besides "Cherenkov" deceleration, a superluminal cosmic 
background radiation may exist
and generate (at much higher energies?) its own GZK 
cutoffs for the superluminal sectors.
\vskip 4mm
To date, there is no well established interpretation [5 - 7] of the
highest-energy cosmic ray events [8].
Primaries (ordinary or superluminal)
originating from superluminal particles are acceptable candidates 
and can possibly escape several problems 
(event configuration, source location, energy dependence...)
faced by cosmic rays produced at ordinary sources.
\vskip 6mm
{\sf Note.} Previous papers on the subject are references [9] - [12].
\vskip 6mm
{\bf References}
\vskip 4mm
\noindent
[1] See, for instance, S.S. Schweber, "An 
Introduction to Relativistic Quantum Field Theory". Row, Peterson and Co. 1961 .
\par
\noindent
[2] See, for instance, "Tachyons, Monopoles 
and Related Topics", Ed. E. Recami. North-Holland 1978 .
\par
\noindent
[3] See, for instance, E. Recami in [2].
\par 
\noindent
[4] See, for instance, P. West, Introduction to 
Supersymmetry and Supergravity, World Scientific 1990 .
\par
\noindent
[5] See, for instance, the Proceedings of TAUP95 , 
Nucl. Phys. B Proc. Suppl. 48 (1996).
\par
\noindent
[6] K. Greisen, Phys. Rev. Lett. 16 , 748 (1966); G.T. Zatsepin and
V. A. Kuzmin, Pisma Zh. Eksp. Teor. Fiz. 4 , 114 (1966).
\par
\noindent
[7] F. W. Stecker, Phys.~Rev.~180~,~1264~(1969);
F. Halzen, R.A. Vazquez, T. Stanev and V.P. Vankov, 
Astropart.~Phys.~3~,~151~(1995);
T.J. Weiler and T.W. Kephart, paper presented at the Dark Matter'96 
Conference, UCLA February 14-16 1996 ,
astro-ph/9605156~.
\par
\noindent
[8] S. Yoshida et al.,~Astropart.~Phys.~3~,~105 (1995);
D.J. Bird et al. Astrophys. J. 424~, 491 (1994) and 
Astrophys. J. 441~, 144 (1995);
M.A. Lawrence, R.J.O. Reid and A.A. Watson, J. Phys. G17 , 733 (1991);
B.N. Afanasiev et al. in Proceedings of the XXIV International 
Cosmic Ray Conference, Rome 1995 , Vol. 2 ,
page 756.
\par 
\noindent 
[9] L. Gonzalez-Mestres, "Properties of a possible class of particles
able to travel faster than light", Proceedings of the
Moriond Workshop on "Dark Matter in Cosmology, Clocks and Tests of 
Fundamental Laws", Villars (Switzerland), January 21 - 28 1995 ,
Ed. Fronti\`eres. Paper astro-ph/9505117 of electronic library.
\par   
\noindent
[10] L. Gonzalez-Mestres, "Cosmological implications of a possible
class of particles able to travel faster than light",
Proceedings of the IV International Conference on Theoretical and 
Phenomenological Aspects of Underground Physics,
TAUP95 , Toledo (Spain) September 1995~, Ed. Nuclear Physics Proceedings. 
Paper astro-ph/9601090 of electronic library.
\par
\noindent
[11] L. Gonzalez-Mestres, "Superluminal Particles 
and High-Energy Cosmic Rays", May 1996~. Paper astro-ph/9606054 
of electronic library.
\par
\noindent
[12] L. Gonzalez-Mestres, "Physics, cosmology and experimental signatures of a
possible new class of superluminal particles", to be published in the
Proceedings of the International Workshop on the Identification of Dark Matter,
Sheffield (England, United Kingdom), September 1996 .
Paper astro-ph/9610089 of
electronic library.
\end{document}